\documentstyle[here,epsfig]{mn2e}
\def\simlt{\mathrel{\rlap{\lower 3pt\hbox{$\sim$}}\raise 2.0pt\hbox{$<$}}}
\def\simgt{\mathrel{\rlap{\lower 3pt\hbox{$\sim$}} \raise 2.0pt\hbox{$>$}}}

\def\gtsima{$\; \buildrel > \over \sim \;$}
\def\ltsima{$\; \buildrel < \over \sim \;$}
\def\gtrsim{\lower.5ex\hbox{\gtsima}}
\def\lesssim{\lower.5ex\hbox{\ltsima}}
\def\url#1{{\ttfamily\def\/{/\diskretionary{}{}{}}#1}}

\newcommand{\q}{\begin{equation}}
\newcommand{\qa}{\begin{eqnarray}}
\newcommand{\qs}{\begin{eqnarray*}}
\newcommand{\nq}{\end{equation}}
\newcommand{\nqa}{\end{eqnarray}}
\newcommand{\nqs}{\end{eqnarray*}}

\begin{document}

%\title[The fate of ring galaxies] 
%{The fate of ring galaxies}
%\title[Are ring galaxies the progenitors of GLSBs?]{Are ring galaxies the progenitors of giant low surface brightness galaxies?}
\title[Lopsided galaxies]{Lopsided galaxies: the case of NGC~891}
\author[Mapelli et al.]
{M. Mapelli$^{1}$, B. Moore$^{1}$, J. Bland-Hawthorn$^{2}$
%{M. Mapelli$^{1}$, R. Sancisi$^{2,3}$, B. Moore$^{1}$
\\
$^{1}$ Institute for Theoretical Physics, University of Z\"urich, Winterthurerstrasse 190, CH-8057, Z\"urich, Switzerland; {\tt mapelli@physik.unizh.ch}\\
$^{2}$ Institute of Astronomy, School of Physics, University of Sydney,  Australia
%Kapteyn Astronomical Institute, University of Groningen, Postbus 800, 9747 AD Groningen, the Netherlands\\
%$^{3}$ INAF - Astronomical Observatory, via Ranzani 1, 40127, Bologna, Italy\\
}

\maketitle \vspace {7cm }

\begin{abstract}
It has been known for a long time that a large fraction of disc galaxies are lopsided. We simulate three different mechanisms that can induce lopsidedness: flyby interactions, gas accretion from cosmological filaments and ram pressure from the intergalactic medium. Comparing the morphologies, HI spectrum, kinematics and $m=1$ Fourier components, we find that all of these mechanisms can induce lopsidedness in galaxies, although in different degrees and with observable consequences. The timescale over which lopsidedness persists suggests that flybys can contribute to $\sim{}20$ per cent of lopsided galaxies.
We focus our detailed comparison on the case of NGC~891, a lopsided, edge-on galaxy with a nearby companion (UGC~1807). We find that the main properties of NGC~891 (morphology, HI spectrum, rotation curve, existence of a gaseous filament pointing towards UGC~1807) favour a flyby event for the origin of lopsidedness in this galaxy.
%, even if gas accretion and ram pressure cannot be excluded.
% can be reproduced by a flyby interaction better than by gas accretion and ram pressure. 
\end{abstract}

\section{Introduction}
A high fraction of disc galaxies are lopsided, i.e. their gas and/or stellar component extend further out on one side of the galaxy than on the other (Baldwin, Lynden-Bell \&{} Sancisi 1980; Block et al. 1994; Richter \&{} Sancisi 1994; Rix \& Zaritsky 1995; Schoenmakers, Franx \& de Zeeuw 1997; Zaritsky \& Rix 1997; Matthews, van Driel \& Gallagher 1998; Haynes et al. 1998; Swaters et al. 1999; Bournaud et al. 2005, hereafter B05; see Sancisi et al. 2008 for a review). 

The gaseous component of the disc is particularly affected by this phenomenon. Richter \& Sancisi (1994) show that the lopsidedness of a galaxy can be inferred from asymmetries in its global HI profile, and estimate, from the analysis of 1700 HI spectra, that $\gtrsim{}50$ per cent of disc galaxies are lopsided in the gaseous component.
Haynes et al. (1998) confirm this result by the analysis of high signal-to-noise HI spectra of 104 galaxies, and suggest that some of the asymmetries may be induced by nearby companions (e.g. NGC~5324). Matthews et al. (1998) indicate that the incidence of gas lopsidedness is higher in the late-type galaxies ($\sim{}77$ per cent).
The kinematics of the gas is often affected by lopsidedness: Swaters et al. (1999) find that the rotation curve of lopsided galaxies is rising more steeply on one side than on the other.

 Rix \& Zaritsky (1995) and Zaritsky \& Rix (1997), using near-infrared photometry of nearly face-on spiral galaxies, show that even the stellar component is lopsided in $\sim{}30$ per cent of their sample. Similarly, Rudnick \&{} Rix (1998), using {\it R}$-$band photometry, find that $\sim{}20$ per cent of their sample of nearly face-on early-type disc galaxies (S0 to Sab) is lopsided in the stellar component. Thus, the incidence of stellar lopsidedness is similar for late-type and for early-type disc galaxies, although slightly lower in the latter case. The analysis of 25155 lopsided galaxies from the Sloan Digital Sky Survey (Reichard et al. 2008) confirms that the lopsided distribution of stellar light is due to a corresponding lopsidedness in the stellar mass. Finally, images and spatially integrated spectra of late-type galaxies (Rudnick, Rix \& Kennicutt 2000) suggest a correlation between star formation and lopsidedness.

The hypothesis that lopsidedness is due to galaxy interactions has been long discussed.
Based on optical images, Odewahn (1994) finds that 71 of 75 lopsided Magellanic spirals have a nearby companion. However, Wilcots \& Prescott (2004) obtain HI data of 13 galaxies from Odewahn (1994) and show that only four of them have HI-detected neighbours.
%two of them have currently interacting companions. 
Thus, either lopsidedness is not related to galaxy interactions, or the asymmetries produced by these interactions are long-lived (surviving for $\gtrsim{}2$ orbital times after the encounter) and the lopsidedness persists even when the companion is quite far-off.

From the theoretical point of view, the N-body simulations by Walker, Mihos \& Hernquist (1996) suggest that minor mergers can induce lopsidedness over a long timescale ($\gtrsim{}$ 1 Gyr). However, B05 indicate that 
the lopsidedness produced by minor mergers disappears when the companion is completely disrupted. Since most of observed lopsided galaxies are not undergoing mergers, the minor-merger scenario does not seem viable. 
B05 indicate that the most likely mechanism to produce lopsidedness is the accretion of gas from cosmological filaments.
Alternative models suggest that baryonic lopsidedness can be induced by a lopsided dark matter halo (Jog 1997, 2002; Angiras et al. 2007) or by the fact that the disc is off-centre with respect to the dark matter halo (Levine \& Sparke 1998; Noordermeer, Sparke \& Levine 2001). 
 
In this paper, we address the problem of the origin of lopsidedness by means of N-body/smooth particle hydrodynamics (SPH) simulations. In particular, we re-analyze in more detail the hypothesis of gas accretion, already proposed by B05, and we consider two new possible scenarios: the role of flyby interactions with smaller companions and that of ram pressure from the intergalactic medium (IGM).

 For a comparison with observational data, we focus on the case of the edge-on galaxy NGC~891. We stress that quantifying lopsidedness in edge-on galaxies is more difficult than in face-on galaxies, as bright regions on one side of the disc can be confused with lopsidedness. However, the lopsidedness of  NGC~891 is well assessed (Sancisi \& Allen 1979; Baldwin et al. 1980; Rupen 1991; Swaters, Sancisi \& van der Hulst 1997). 
%NGC~891. This almost edge-on galaxy is one of the prototypes of lopsided galaxies (Sancisi \& Allen 1979; Baldwin et al. 1980; Rupen 1991; Swaters, Sancisi \& van der Hulst 1997). {\bf We stress that quantifying lopsidedness in edge-on galaxies is more difficult than in face-on ones, as bright regions on one side can be taken for lopsidedness. However, the lopsidedness in  NGC~891 } 
Furthermore, for NGC~891 recent HI observations are available, among the deepest ever obtained for an external galaxy (Oosterloo, Fraternali \& Sancisi 2007, hereafter O07). This galaxy also shows many interesting peculiarities, e.g. the existence of a gaseous filament extending up to $\sim{}20$ kpc vertically from the disc and located at $\sim{}10$ kpc from the centre of the galaxy. Finally, NGC~891 has also a smaller, gas-rich companion, UGC~1807, located at a projected distance of $\sim{}80$ kpc, in the direction of the above mentioned gaseous filament.

\section{Models and simulations}
In this paper, we simulate three different processes: i) flyby interactions; ii) accretion from gaseous filaments; iii) ram pressure from the IGM.
For all these scenarios we use a galaxy model similar to NGC~891 and whose main properties are listed in Table~1.
Such galaxy model has been generated by using the method already described in Mapelli 2007 (hereafter M07; see also Hernquist 1993; Mapelli, Ferrara \& Rea 2006; Mapelli et al. 2008a, 2008b). Here we briefly summarize the most important points, referring to M07 for the details.
The galaxy model has four different components:

\begin{itemize}
\item[-] a Navarro, Frenk \& White (1996, NFW) dark matter halo with  virial mass $M_{vir}=1.4\times{}10^{11}\,{}M_\odot{}$ (O07),  virial radius $R_{200}=104$ kpc, and  concentration $c=12$;

\item[-] a stellar exponential Hernquist disc (Hernquist 1993; M07) with mass $M_d=10^{10}\,{}M_\odot{}$, scalelength $R_d=4.4$ kpc (Shaw \& Gilmore 1989), and scaleheight $z_0=0.1\,{}R_d$;

\item[-] a stellar spherical Hernquist bulge (Hernquist 1993; M07) with mass $M_b=2\times{}10^{9}\,{}M_\odot{}$, and scalelength $a=0.1\,{}R_d$;

\item[-] a gaseous exponential Hernquist disc (Hernquist 1993; M07) with mass $M_g=4.1\times{}10^{9}\,{}M_\odot{}$ (O07), scalelength $R_g=R_d$, and scaleheight $z_g=0.1\,{}R_g$. The gas is allowed to cool down to a temperature of $10^4$~K, and to form stars according to the Schmidt law (Katz 1992).

\end{itemize}

The model of NGC~891 has $619\,{}500$ halo dark matter particles, $500\,{}000$ stellar disc particles, $100\,{}000$ bulge particles and $205\,{}000$ gaseous disc particles.
Dark matter particles have mass equal to $2.23\times{}10^{5}\,{}M_\odot{}$, whereas disc, bulge and gas particles have mass equal to $2.23\times{}10^4\,{}M_\odot{}$. Softening lengths are 0.2 kpc for halo particles and 0.1 kpc for disc, bulge and gas particles (for the criterion used to estimate the softening, see Dehnen 2001). The initial smoothing length of the gas is $\sim{}0.1$ kpc.

The simulations have been carried out with the parallel $N-$body/SPH code GASOLINE (August 2005 version; Wadsley, Stadel \& Quinn 2004) on the cluster {\it zbox}2\footnote{{\tt http://www-theorie.physik.unizh.ch/\~{}dpotter/zbox/}} at the University of Z\"urich. 

%%%%%%%%%%%%%%%%%%%%%%%%%%%%%%%%%%% TABLE 1 %%%%%%%%%%%%%%%%%%%%%%%%%%%%%%%%%%%
\begin{table}
\begin{center}
\caption{Initial parameters for the model of NGC~891 and of UGC~1807.}
\begin{tabular}{ccc}
\hline
%\vspace{0.1cm}
          & NGC~891 & UGC~1807\\
\hline
%\vspace{0.1cm}
$M_{vir}/M_\odot{}$ & $1.4\times{}10^{11}$   & $7.65\times{}10^{9}$\\
$R_{200}/$kpc       & 104                    & 41\\
$M_d/M_\odot{}$     & $10^{10}\,{}M_\odot{}$ & $9\times{}10^{8}$\\
$R_d/$kpc           & 4.4                    & 0.5\\
$z_d$               & 0.1 $R_d$              & 0.1 $R_d$\\
$M_b/M_\odot{}$     & $2\times{}10^{9}$      & 0 \\
$a$                 & 0.1 $R_d$              & 0\\
$M_g/M_\odot{}$     & $4.1\times{}10^{9}$    & $4.5\times{}10^{8}$\\
$R_g$               & $R_d$                  & $R_d$\\
$z_g$               & 0.1 $R_g$              & 0.1 $R_g$\\
\hline
\end{tabular}
\end{center}
%\begin{flushleft}
%{\footnotesize $^{a}$The values adopted here for $R_d$ are about twice (for runs A and B) and thrice (for run C) the value predicted by Mo, Mao \& White (1998).}\\
%\end{flushleft}
\label{tab_1}
\end{table}
%%%%%%%%%%%%%%%%%%%%%%%%%%%%%%%%%%%%%%%%%%%%%%%%%%%%%%%%%%%%%%%%%%%%%%%%%%%%%%%
\subsection{Flyby interaction}
In the case of flyby interactions, we simulate an intruder galaxy with the properties of UGC~1807 (see Table~1), i.e.

\begin{itemize}
\item[-] a NFW dark matter halo with  virial mass $M_{vir}=7.65\times{}10^{9}\,{}M_\odot{}$ (O07),  virial radius $R_{200}=41$ kpc, and  concentration $c=12$;

\item[-] a stellar exponential Hernquist disc with mass $M_{d}=9\times{}10^{8}\,{}M_\odot{}$, scalelength $R_{d}=0.5$ kpc, and scaleheight $z_{0}=0.1\,{}R_{d}$;

\item[-] a gaseous exponential Hernquist disc with mass $M_{g}=4.5\times{}10^{8}\,{}M_\odot{}$ (O07), scalelength $R_{g}=R_{d}$, and scaleheight $z_{g}=0.1\,{}R_{g}$. The recipes for gas cooling and star formation are the same as for the target galaxy.
\end{itemize}

As the mass of dark matter and baryonic particles in the intruder are $2.23\times{}10^{5}M_\odot{}$ and  $2.23\times{}10^{4}M_\odot{}$, respectively (i.e. the same as in the target galaxy),  the intruder is initially composed by $34\,{}300$ dark matter particles, $40\,{}300$ stellar disc particles and $20\,{}200$ gas particles. Also the softening lengths are the same as in the target.

We made various check runs, in order to find the initial centre-of-mass position and velocity of the intruder for which its final position and velocity best match the observations of UGC~1807. This best match is achieved for an initial centre-of mass position ({\it x}, {\it y}, {\it z})=($-$38, 38, 80) kpc, and for an  initial centre-of mass velocity ({\it v$_x$}, {\it v$_y$}, {\it v$_z$})=(130, $-$130, 200) km s$^{-1}$.

\subsection{Gas accretion from filaments}
The existence of filaments of cold gas that accrete onto galaxies is predicted by cosmological SPH simulations (Katz \& White 1993; Katz et al. 1994; Keres et al. 2005; Ocvirk, Pichon \&{} Teyssier 2008). It has also been proposed as a mechanism for the origin of polar ring galaxies (Macci\`o et al 2005).
B05 showed that gas accretion might induce lopsidedness, similar to the one observed in isolated spiral galaxies such as NGC~1637.

We set up our simulations in a similar way to  B05, i.e. we model a cylindrical, uniform gas filament which accretes onto the galaxy with a given accretion rate $\dot{M}_\odot{}$ ($\lesssim{}6\,{}M_\odot{}$ yr$^{-1}$). The filament is corotating with the galaxy. Different from B05, who consider only coplanar filaments, we also simulate the case of non-coplanar accretion. In the following, we will present the results for three different runs in which gas filaments are simulated (see Table 2). In all of them the filament has a radius of 12.5 kpc, a relative velocity  of infalling onto the galaxy $v_{infall}=100$ km s$^{-1}$, an accretion rate $\dot{M}_\odot{}=6\,{}M_\odot{}$ yr$^{-1}$, and a 25 kpc off-set with respect to the centre of the galaxy (see fig. 16 of B05).

In run ACC1 the filament is coplanar to the galaxy (see fig. 16 of B05).
%({\it x}, {\it y}, {\it z})=(-25, 250, 0) kpc
%({\it v$_x$}, {\it v$_y$}, {\it v$_z$})=(0, -100, 0) km s$^{-1}$
In run ACC2 the filament is still parallel to the plane of the galaxy, but is shifted of 10 kpc above it. 
Finally, in run ACC3 the filament has an inclination of $-20^\circ{}$ with respect to the plane of the galaxy.
The existence of non-coplanar filaments, and in particular  of the case ACC3, is supported by cosmological simulations (Keres et al. 2005).

The total number of particles in the filament in the above runs is $543\,{}910$ and the mass of each particle is $2.23\times{}10^4\,{}M_\odot{}$. We made also other check runs with smaller  $\dot{M}_\odot{}$ (down to $2\,{}M_\odot{}$ yr$^{-1}$). We integrate each run for $\sim{}$1 Gyr.

%%%%%%%%%%%%%%%%%%%%%%%%%%%%%%%%%%% TABLE 2 %%%%%%%%%%%%%%%%%%%%%%%%%%%%%%%%%%%
\begin{table}
\begin{center}
\caption{Parameters of runs of gas accretion from filaments}
\begin{tabular}{cccc}
\hline
%\vspace{0.1cm}
run   & coplanar & $z$-shift$^{a}$ & inclination$^b$\\
\hline
%\vspace{0.1cm}
ACC1  &   Yes     &    0      &   0\\
ACC2  &   No      &   10 kpc  &   0\\
ACC3  &   No      &    0      &   $-20^\circ{}$\\
\hline
\end{tabular}
\end{center}
\begin{flushleft}
{\footnotesize $^{a}$Off-set of the filament along the $z$-axis with respect to the plane of the galaxy.\\
\footnotesize $^{a}$Inclination of the filament with respect to the plane of the galaxy.\\}
\end{flushleft}
\label{tab_2}
\end{table}
%%%%%%%%%%%%%%%%%%%%%%%%%%%%%%%%%%%%%%%%%%%%%%%%%%%%%%%%%%%%%%%%%%%%%%%%%%%%%%%

\subsection{Ram pressure from the IGM}
The characteristic signatures of ram pressure have recently been observed also in dwarf irregular galaxies of the Local Group (McConnachie et al. 2007). This suggests the existence of a non-negligible amount of diffuse IGM even in poor groups.
In particular, gas is ram-pressure stripped from a galaxy if the density of the IGM is (Gunn \& Gott 1972; McConnachie et al. 2007)
\begin{equation}\label{eq:ram}
n_{\rm IGM}\gtrsim{}3.7\times{}10^{-6}\textrm{ cm}^{-3}\,{}\left(\frac{100\textrm{ km s}^{-1}}{v_{rel}}\right)^2\,{}\left(\frac{\Sigma{}_{\rm HI}}{10^{19}\textrm{ cm}^{-2}}\right)^2,
\end{equation}
where $v_{rel}$ is the relative velocity between the galaxy and the IGM and $\Sigma{}_{\rm HI}$ is the  column density of HI.

Thus, if the IGM density is higher than the threshold value in equation~(\ref{eq:ram}), even galaxies which are isolated or in small groups can suffer ram pressure and the distribution of their gas might appear lopsided. 

According to Bland-Hawthorn, Freeman \& Quinn (1997), ram pressure heating can explain the fact that some galaxies (e.g. the lopsided galaxy NGC 253) have a truncated HI disc but present high temperature, ionized gas extending beyond the HI disc. Similarly, Ryder et al. (1997) suggest that ram pressure is responsible for the HI 'wake' observed in the spiral galaxy NGC~7421. 
%{\bf Finally, Rand, Wood \& Benjamin (2008) find that photo-ionization due to massive stars in the disc of NGC~891 is not sufficient to explain the ionization level of its gaseous halo.}
%The high temperature of the external gas is indicated by the enhanced NII/Ha ratio.  
%suggested as responsible for the 

To check the hypothesis that lopsidedness is induced by ram pressure, we made a simulation where the galaxy is moving into a cylinder of uniform gas with density $n_{\rm IGM}=5\times{}10^{-5}$ cm$^{-3}$ and $v_{rel}=200$ km s$^{-1}$. We made different runs, changing the inclination of the relative velocity $v_{rel}$ with respect to the plane of the galaxy between 0 and 45$^{\circ}$.
The total number of IGM particles in the above runs is $2\,{}088\,{}630$ and the mass of each particle is $2.23\times{}10^4\,{}M_\odot{}$. We integrate each run for $\sim{}$1.5 Gyr.
%\section{Simulations}

%%%%%%%%%%%%%%%%%%%%%%%%%%%%%%%%%%% FIGURE 1 %%%%%%%%%%%%%%%%%%%%%%%%%%%%%%%%%%
\begin{figure}
\center{{
\epsfig{figure=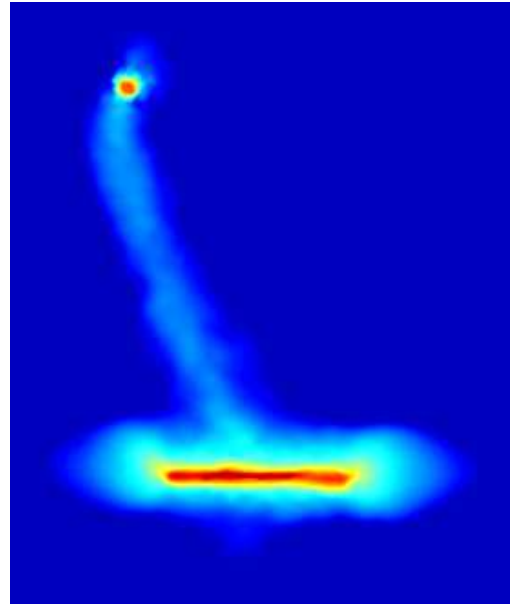,height=8cm}
}}
\caption{\label{fig:fig1} Density map of gas, projected along the $y$-axis, for the target galaxy (edge-on) and the companion (face-on), at $t=640$ Myr after the beginning of the simulation and $\sim{}300$ Myr after the flyby. The frame measures 105 and 129 kpc, along the short and the long side, respectively. The density goes from $2.23\times{}10^{-3}$ (blue on the web) to $2.23\times{}10^1\,{}M_\odot{}$ pc$^{-2}$ (red on the web) in logarithmic scale. 
%The frame measures 105$\times{}$129 kpc.
}
\end{figure}
%%%%%%%%%%%%%%%%%%%%%%%%%%%%%%%%%%%%%%%%%%%%%%%%%%%%%%%%%%%%%%%%%%%%%%%%%%%%%%%

\section{Results}
The main result of this paper is that all the considered processes induce lopsidedness in the gaseous component. In the following, we discuss the details and the differences among the three scenarios.
We firstly describe the morphological features of the simulated gaseous and (global and young) stellar component in the three scenarios (Sections~\ref{sec:fly}-\ref{sec:ram}). Then, we provide a more quantitative estimate of some observational quantities, i.e. the Fourier $m=1$ component (Section~\ref{sec:m1}), the HI spectrum and the  rotation curve (Section~\ref{sec:kine}).
 
\subsection{Flybys}\label{sec:fly}
Fig.~\ref{fig:fig1} shows the density of the gas in the target galaxy (seen edge-on) and in the companion galaxy (face-on) $\sim{}300$ Myr after the interaction, when the projected distance between the two galaxies is $\sim{}80$ kpc, i.e. comparable with the observations of NGC~891 and UGC~1807.

The simulation matches some of the properties of the NGC~891$-$UGC~1807 system. First, a gaseous filament connects the two galaxies in the simulation. The filament starts at $\sim{}$10 kpc (projected distance along the $y$-axis) from the centre of the target galaxy, which is the position where the intruder reached the closest approach with the target. This filament is mostly due to the gas stripped from the companion galaxy. 
The position of the filament is the same as in the observations (O07). However, in the simulation the density of gas is more or less constant along the filament, whereas in the data only the part of the filament which is closest to NGC~891 is visible.

%%%%%%%%%%%%%%%%%%%%%%%%%%%%%%%%%%% FIGURE 2 %%%%%%%%%%%%%%%%%%%%%%%%%%%%%%%%%%
%\begin{figure*}
\begin{figure}
\center{{
\epsfig{figure=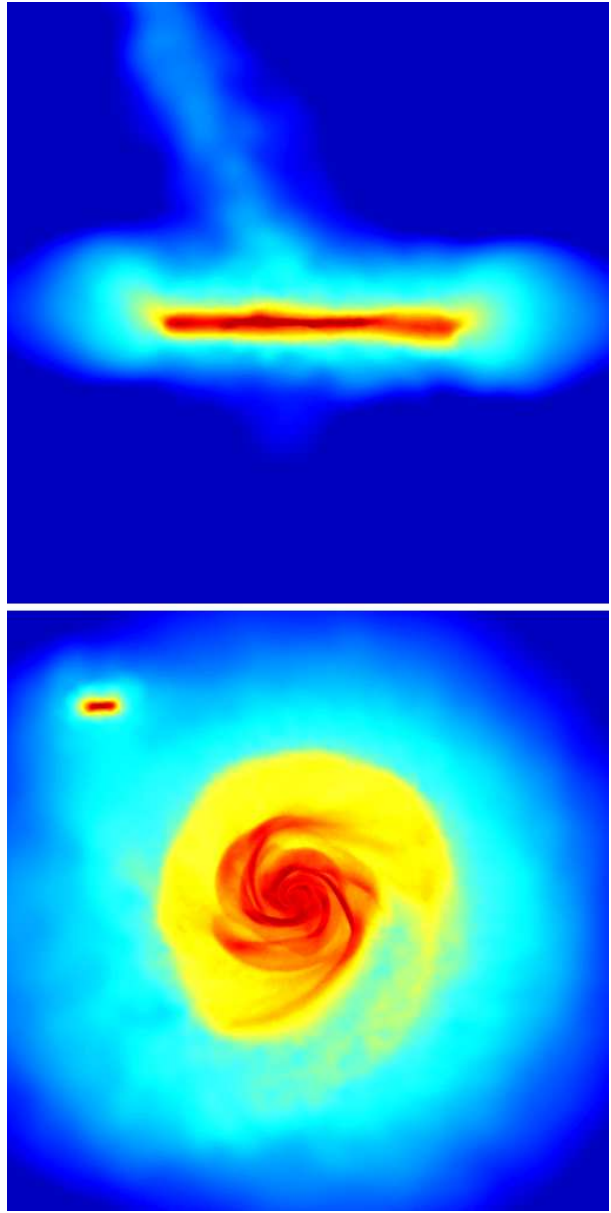, width=8.0cm}
}}
\caption{\label{fig:fig2} Density map of gas in the target galaxy, at $t=640$ Myr after the beginning of the simulation and $\sim{}300$ Myr after the flyby. Top panel: the target galaxy is seen edge-on and the density is projected along the  $y$-axis (this panel is a zoom of Fig.~\ref{fig:fig1}). Bottom panel: the target galaxy is seen face-on  and the density is projected along the  $z$-axis. The frames measure both 80 kpc per edge. The density goes from $2.23\times{}10^{-3}$ to $2.23\times{}10^1\,{}M_\odot{}$~pc$^{-2}$ in logarithmic scale. 
}
\end{figure}
%\end{figure*}
%%%%%%%%%%%%%%%%%%%%%%%%%%%%%%%%%%%%%%%%%%%%%%%%%%%%%%%%%%%%%%%%%%%%%%%%%%%%%%%
%%%%%%%%%%%%%%%%%%%%%%%%%%%%%%%%%%% FIGURE 3 %%%%%%%%%%%%%%%%%%%%%%%%%%%%%%%%%%
\begin{figure}
\center{{
\epsfig{figure=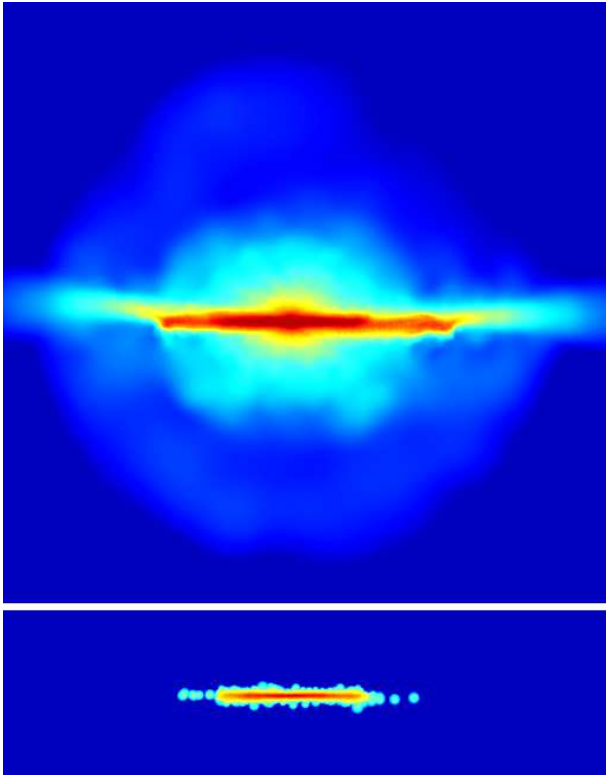,width=8cm}
}}
\caption{\label{fig:fig3}  Density map of stars, projected along the $y$-axis, in the target galaxy, at $t=640$ Myr after the beginning of the simulation and $\sim{}300$ Myr after the flyby. Top panel: all stars of the target galaxy. The frame measures 80 kpc per edge and the density goes from $2.23\times{}10^{-3}$ to $2.23\times{}10^2\,{}M_\odot{}$~pc$^{-2}$  in logarithmic scale. Bottom panel: young stars ($\le{}100$ Myr) of the target galaxy. The frame measures  80 and 22 kpc, along the long and the short side, respectively. The density goes from $7.05\times{}10^{-6}$ to $2.23\times{}10^2\,{}M_\odot{}$~pc$^{-2}$  in logarithmic scale. 
}
\end{figure}
%%%%%%%%%%%%%%%%%%%%%%%%%%%%%%%%%%%%%%%%%%%%%%%%%%%%%%%%%%%%%%%%%%%%%%%%%%%%%%%

Second, the edge-on disc of the target galaxy is clearly  lopsided (see the zoom in Fig.~\ref{fig:fig2}). In particular, the disc extends further out on the side of the galaxy which is opposite with respect to the intruder. Indeed, the lopsidedness is induced by the flyby, as the perturbation originated exactly where the intruder penetrated the disc of the target. At the time shown by Figs.~\ref{fig:fig1} and ~\ref{fig:fig2} (i.e. $\sim{}300$ Myr after the flyby), the galactic disc has completed a half-rotation, and the perturbation appears on the other side of the disc with respect to its initial position.

Third, Fig.~\ref{fig:fig3} shows that even the stellar component of the simulated galaxy has some degree of lopsidedness, although less evident than in the gas. The observations also show that the stellar component of NGC~891 is slightly lopsided,  especially the young population (van der Kruit \& Searle 1981; Rand, Kulkarni \& Hester 1990; Hoopes, Walterbos \& Rand 1999; Kamphuis et al. 2007a; Kamphuis et al. 2007b). We also ran check simulations with different disc scalelength $R_d$ (between 2.2 and 8.8 kpc), but we did not find significant differences from the point of view of lopsidedness.
Then, the flyby scenario seems to reproduce quite well the lopsidedness of NGC~891.

Furthermore, the simulations suggest that the perturbation induced by flybys is long-lived ($\gtrsim{}500$ Myr), explaining the origin of lopsidedness even in galaxies which have no longer interacting companions. This might be important, as many studies do not find any correlation between lopsidedness and nearby companions (Rix \& Zaritsky 1995; Zaritsky \& Rix 1997; Wilcots \& Prescott 2004; B05).

%persistence of the perturbation $\gtrsim{}300$ Myr after the interaction (i.e. the time which takes the intruder to go at the observed distance from NGC~891) indicates that lopsidedness induced by flybys might be long-lived, and explain

\subsection{Gas accretion}\label{sec:accr}
%%%%%%%%%%%%%%%%%%%%%%%%%%%%%%%%%%% FIGURE 4 %%%%%%%%%%%%%%%%%%%%%%%%%%%%%%%%%%
%\begin{figure*}
\begin{figure}
\center{{
\epsfig{figure=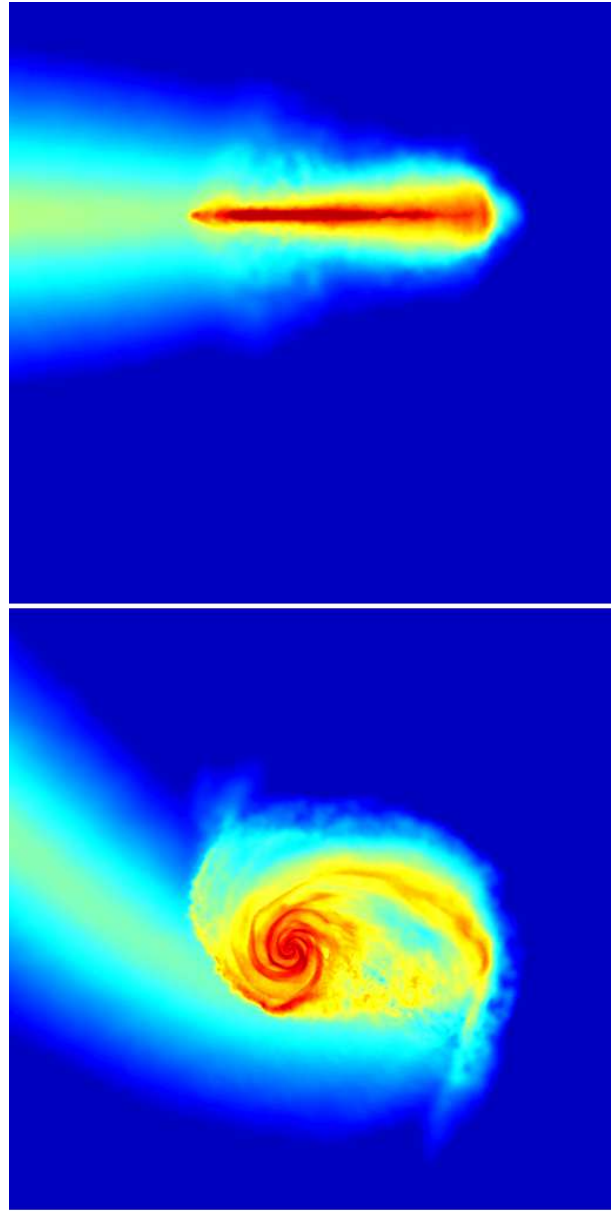,width=8.0cm}
}}
\caption{\label{fig:fig4} 
Density map of gas in run ACC1, at $t=720$ Myr after the beginning of the simulation. Top panel: the galaxy is seen edge-on and the density is projected along the  $y$-axis. Bottom panel: the galaxy is seen face-on and the density is projected along the  $z$-axis. The frames measure both 100 kpc per edge. The density goes from $7.05\times{}10^{-2}$ to $2.23\times{}10^1\,{}M_\odot{}$~pc$^{-2}$ in logarithmic scale. 
}
\end{figure}
%\end{figure*}
%%%%%%%%%%%%%%%%%%%%%%%%%%%%%%%%%%%%%%%%%%%%%%%%%%%%%%%%%%%%%%%%%%%%%%%%%%%%%%%

%%%%%%%%%%%%%%%%%%%%%%%%%%%%%%%%%%% FIGURE 5 %%%%%%%%%%%%%%%%%%%%%%%%%%%%%%%%%%
\begin{figure}
\center{{
\epsfig{figure=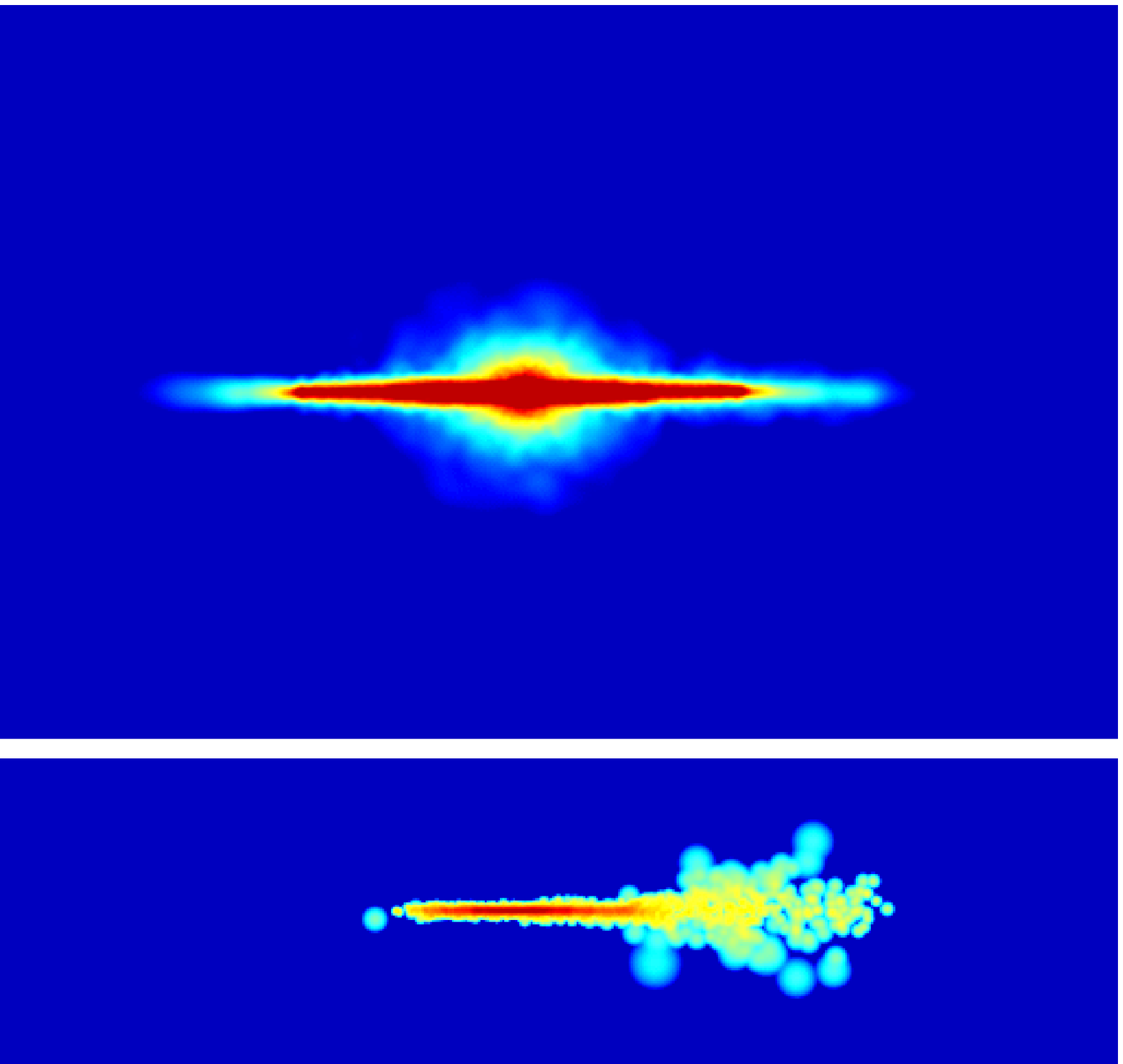,width=8cm}
}}
\caption{\label{fig:fig5} 
Density map of stars in run ACC1, at $t=720$ Myr after the beginning of the simulation. The galaxy is seen edge-on and the density is projected along the  $y$-axis. Top panel: all stars.  The frame measures  100 and 66 kpc, along the long and the short side, respectively. The density goes from $7.05\times{}10^{-2}$ to $2.23\times{}10^1\,{}M_\odot{}$~pc$^{-2}$  in logarithmic scale. Bottom panel: young stars ($\le{}100$ Myr). The frame measures  100 and 27 kpc, along the long and the short side, respectively. The density goes from $7.05\times{}10^{-6}$ to $2.23\times{}10^1\,{}M_\odot{}$~pc$^{-2}$  in logarithmic scale. 
}
\end{figure}
%%%%%%%%%%%%%%%%%%%%%%%%%%%%%%%%%%%%%%%%%%%%%%%%%%%%%%%%%%%%%%%%%%%%%%%%%%%%%%%

Gas accretion from cosmological filaments is also able to produce lopsidedness, as suggested by B05. However, the kind of lopsidedness that might be induced by gas accretion is quite different from the one connected with flybys. This can be seen from the comparison between Fig.~\ref{fig:fig2} and Fig.~\ref{fig:fig4}. Both figures show the density of gas in the edge-on (upper panels) and face-on (lower panels) view of the galaxy, but Fig.~\ref{fig:fig2} refers to the flyby scenario and  Fig.~\ref{fig:fig4} to the gas accretion from a coplanar filament (ACC1). In the case of gas accretion, the lopsidedness appears much more pronounced, especially in the face-on view. Of course, this depends on the chosen accretion rate ($\dot{M}=6\,{}M_\odot{}$ yr$^{-1}$ in ACC1) and on the duration of the accretion phase (720 Myr in Fig.~\ref{fig:fig4}). However, even smaller accretion rates (down to $\dot{M}=2\,{}M_\odot{}$ yr$^{-1}$) and different timescales for the accretion (0.5$-$1.0 Gyr) produce approximately the same morphology.
Therefore, the simulated galaxy looks more like, e.g., NGC~1637 than NGC~891.

Of course, our models suffer from various limitations, such as the intrinsic problems of the SPH treatment of gas (see Agertz et al. 2007), the mass resolution and the fact that we do not account for various feedback processes (e.g. the cooling for temperatures below 10$^4$ K). Thus, we cannot take the morphology inferred from our simulations as a  conclusive proof. However, the morphological difference seen in our simulations might suggest an intrinsic morphological difference in observed galaxies between lopsidedness induced by gas accretion and  lopsidedness induced by flybys. Thus, it would be interesting to compare the morphologies of observed lopsided galaxies, with or without companions, in order to check whether any difference can be found (as an example, we suggest the cases of NGC~1637 and NGC~891).

Furthermore, the total stellar component (top panel of Fig.~\ref{fig:fig5}) does not seem to be lopsided in the gas accretion scenario, at odds with the observations of NGC~891. Instead, the young stellar population (i.e. the simulated star particles which have an age $\le{}100$ Myr) shows a clear lopsidedness (bottom panel of  Fig.~\ref{fig:fig5}). This is likely due to the fact that gas accretion does not affect stars and induces lopsidedness only in the gas component. The distribution of young stars  is also lopsided, because these stars formed after the gas component has become lopsided. We can consider this process a sort of 'star formation' driven lopsidedness (see Section~\ref{sec:m1} for details).

\begin{figure}
\center{{
\epsfig{figure=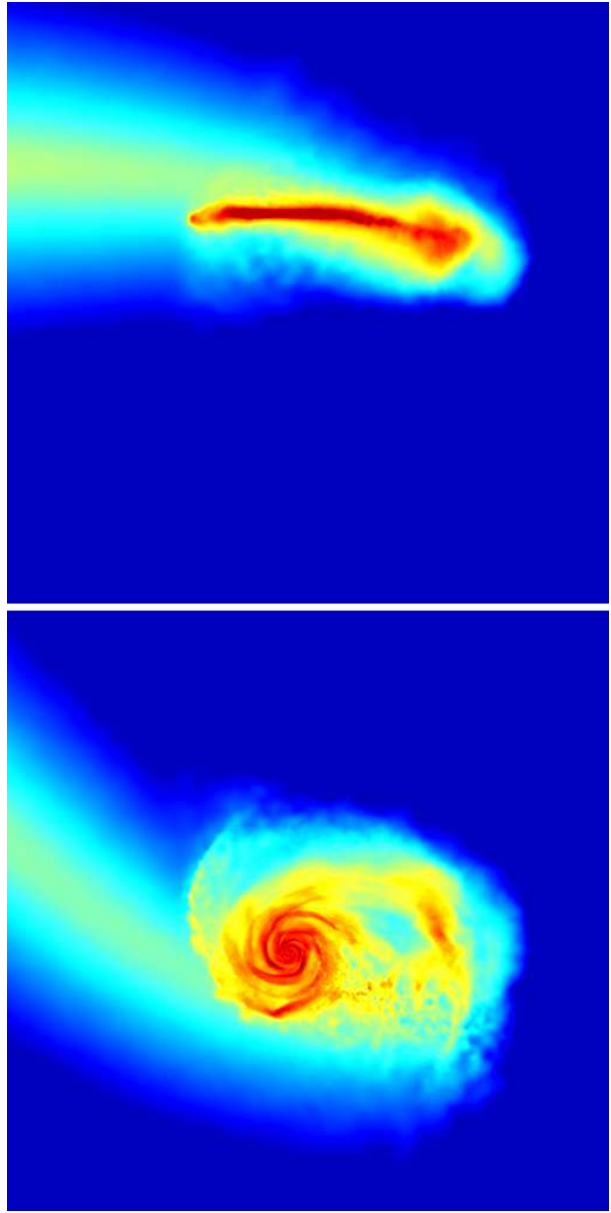,width=8.0cm}
}}
\caption{\label{fig:fig6} 
Density map of gas in run ACC2, at $t=720$ Myr after the beginning of the simulation. Top and bottom panels are the same as in Fig.~\ref{fig:fig4}. The frames measure both 100 kpc per edge. The density scale is the same as in Fig.~\ref{fig:fig4}.}
\end{figure}

\begin{figure}
\center{{
\epsfig{figure=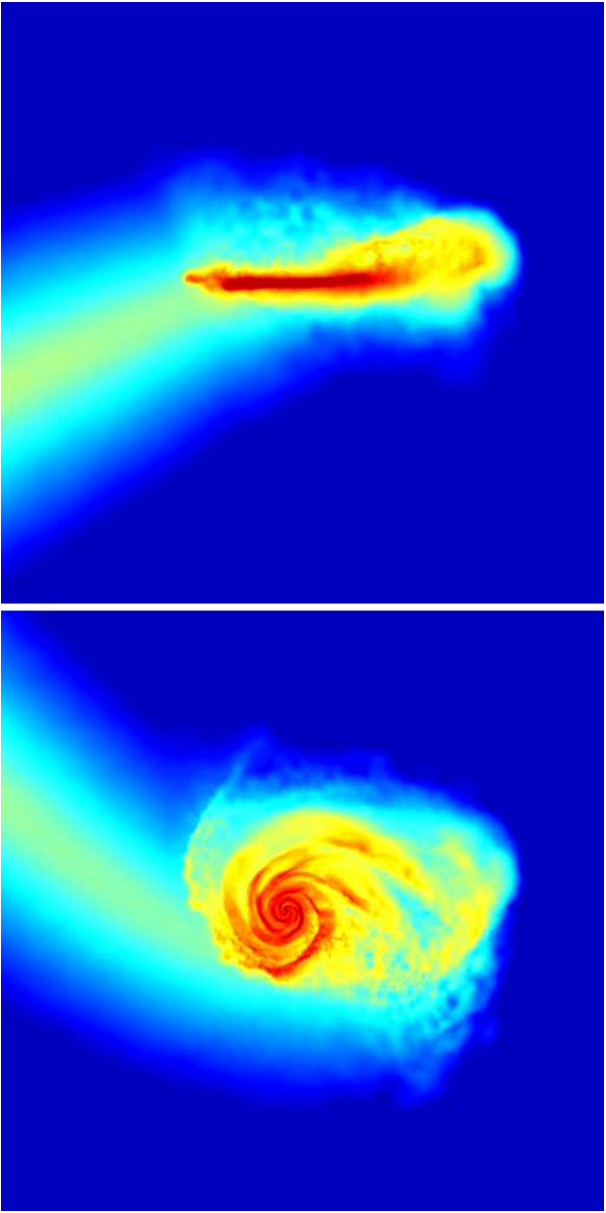,width=8.0cm}
}}
\caption{\label{fig:fig7} 
Density map of gas in run ACC3, at $t=720$ Myr after the beginning of the simulation. Top and bottom panels are the same as in Fig.~\ref{fig:fig4}. The frames measure both 100 kpc per edge. The density scale is the same as in Fig.~\ref{fig:fig4}.}
\end{figure}

The gas accretion from a coplanar filament does not produce anything similar to the gaseous filament observed in NGC~891.
In runs ACC2 and ACC3 we change the orientation of the filament (see Table 2), in order to see whether we can reproduce the filament observed in NGC~891. However, in runs ACC2 and ACC3 (Figs.~\ref{fig:fig6} and \ref{fig:fig7}, respectively) the galaxy disc becomes not only lopsided but also strongly distorted. This feature is not present in NGC~891 and in most of observed lopsided galaxies, whereas it might be somehow connected with warps. Such distortion tends to disappear only if the accretion rate is sufficiently small ($\lesssim{}2\,{}M_\odot{}\textrm{ yr}^{-1}$). As cosmological simulations indicate that non-coplanar filaments exist (Keres et al. 2005), this may put limits on the density and/or on the lifetime of cosmological filaments. Further study of gas filaments in cosmological simulations is required, in order to quantify this effect.
%{\bf In a forthcoming paper we will address this issue in a dedicated way}

On the other hand, non-coplanar gas accretion does not produce any structure like the filament observed in NGC~891, unless we assume that the cosmological filament itself (or its remnant) is still observable today and represents the feature observed in NGC~891.
%%%%%%%%%%%%%%%%%%%%%%%%%%%%%%%%%%% FIGURE 6 %%%%%%%%%%%%%%%%%%%%%%%%%%%%%%%%%%
%\begin{figure*}
%\end{figure*}
%%%%%%%%%%%%%%%%%%%%%%%%%%%%%%%%%%%%%%%%%%%%%%%%%%%%%%%%%%%%%%%%%%%%%%%%%%%%%%%
%%%%%%%%%%%%%%%%%%%%%%%%%%%%%%%%%%% FIGURE 7 %%%%%%%%%%%%%%%%%%%%%%%%%%%%%%%%%%
%\begin{figure*}
%\end{figure*}
%%%%%%%%%%%%%%%%%%%%%%%%%%%%%%%%%%%%%%%%%%%%%%%%%%%%%%%%%%%%%%%%%%%%%%%%%%%%%%%
Indeed, the co-existence of two different cosmological filaments, one coplanar  inducing lopsidedness and the other non-coplanar producing the observed filament in NGC~891, might explain the morphology of this galaxy, but seems quite fine-tuning.
\subsection{Ram pressure}\label{sec:ram}
%%%%%%%%%%%%%%%%%%%%%%%%%%%%%%%%%%% FIGURE 8 %%%%%%%%%%%%%%%%%%%%%%%%%%%%%%%%%%
\begin{figure}
\center{{
\epsfig{figure=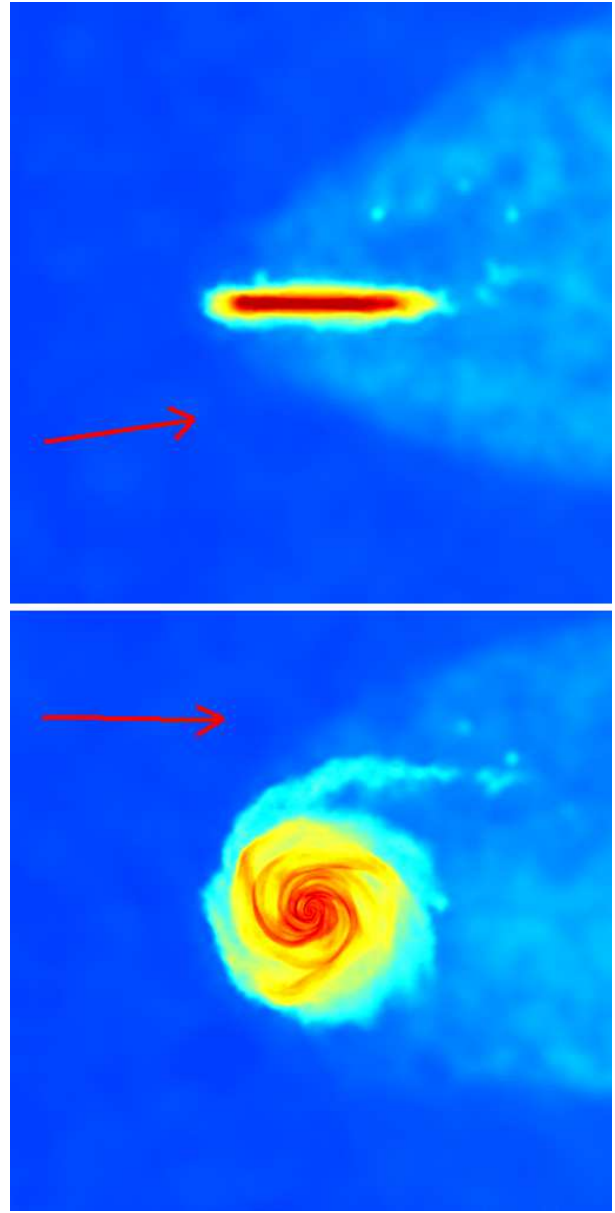,width=8.0cm}
}}
\caption{\label{fig:fig8} 
Density map of gas in the ram pressure scenario, at $t=1$ Gyr after the beginning of the simulation. Top panel: the galaxy is seen edge-on and the density is projected along the  $x$-axis. Bottom panel: the galaxy is seen face-on and the density is projected along the  $z$-axis. The frames measure both 100 kpc per edge. The density goes from $5.60\times{}10^{-2}$ to $2.23\times{}10^1\,{}M_\odot{}$~pc$^{-2}$ in logarithmic scale. The two arrows indicate approximately the direction of the relative velocity between galaxy and IGM (which has an inclination of 10$^\circ{}$ with respect to the plane of the disc).
}
\end{figure}
%%%%%%%%%%%%%%%%%%%%%%%%%%%%%%%%%%%%%%%%%%%%%%%%%%%%%%%%%%%%%%%%%%%%%%%%%%%%%%%
%%%%%%%%%%%%%%%%%%%%%%%%%%%%%%%%%%% FIGURE 8bis %%%%%%%%%%%%%%%%%%%%%%%%%%%%%%%%%%
\begin{figure}
\center{{
\epsfig{figure=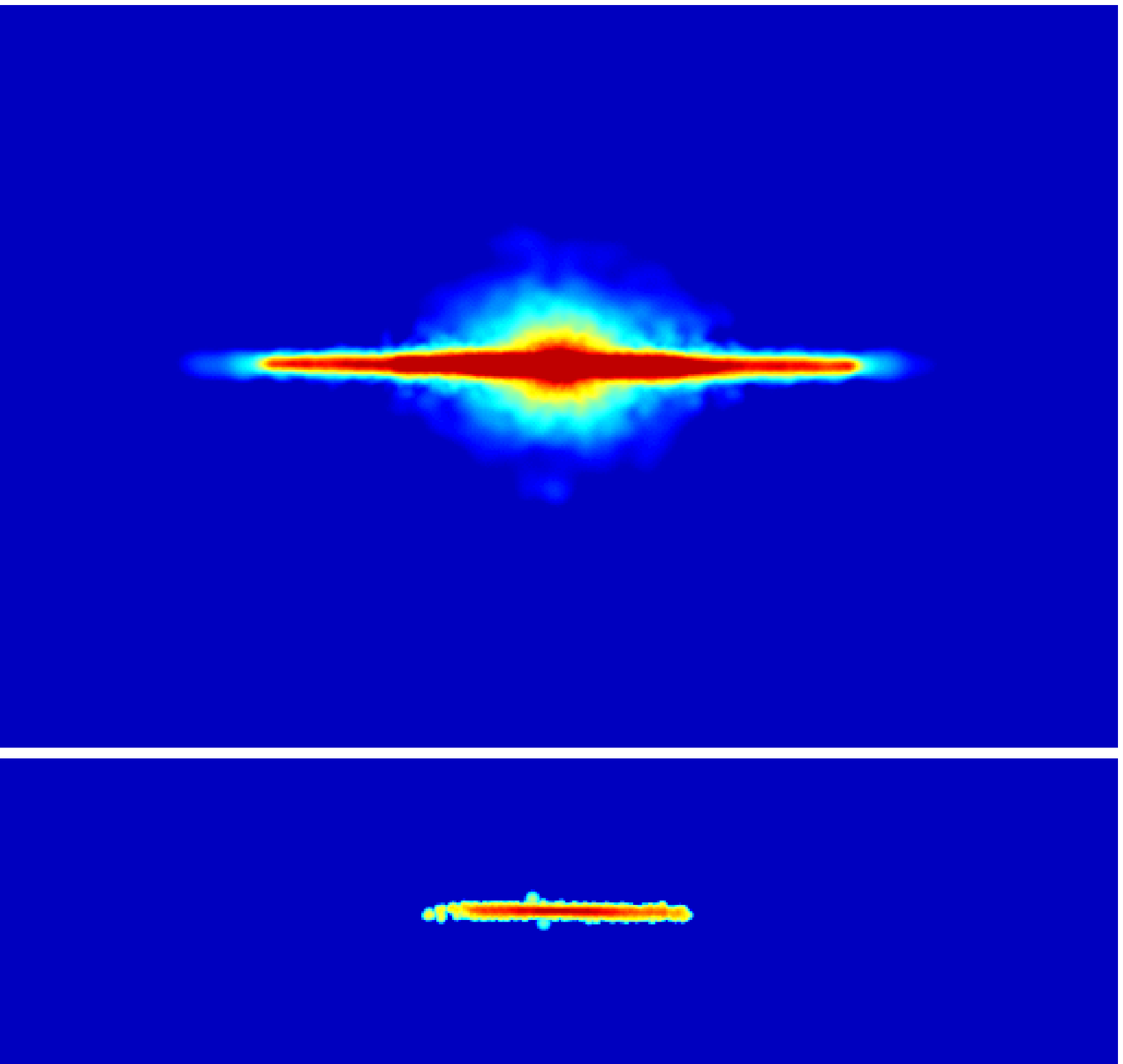,width=8cm}
}}
\caption{\label{fig:fig8bis} 
Density map of stars in the ram pressure scenario, at $t=1$ Gyr after the beginning of the simulation. The galaxy is seen edge-on and the density is projected along the  $x$-axis. 
 Top panel: all stars.  The frame measures  100 and 66 kpc, along the long and the short side, respectively. The density goes from $5.60\times{}10^{-2}$ to $2.23\times{}10^1\,{}M_\odot{}$~pc$^{-2}$  in logarithmic scale. Bottom panel: young stars ($\le{}100$ Myr). The frame measures  100 and 27 kpc, along the long and the short side, respectively. The density goes from $7.05\times{}10^{-6}$ to $2.23\times{}10^1\,{}M_\odot{}$~pc$^{-2}$  in logarithmic scale. 
}
\end{figure}
%%%%%%%%%%%%%%%%%%%%%%%%%%%%%%%%%%%%%%%%%%%%%%%%%%%%%%%%%%%%%%%%%%%%%%%%%%%%%%%
Even ram pressure induces lopsidedness (see Fig.~\ref{fig:fig8}), after a sufficiently large amount of time ($\gtrsim{}600-700$ Myr). However, lopsidedness produced by ram pressure appears less pronounced than that induced by either flybys or gas accretion. This is mainly due to the fact that ram pressure perturbs especially the low density gas in the peripheral regions, but does not affect the central parts of the disc. Thus, the effect of ram pressure is that of producing a faint tidal tail, more than a strong asymmetry. The stellar component, both old and young, does not suffer significant perturbations due to ram pressure (see Fig.~\ref{fig:fig8bis}).

Obviously, this result depends on our initial parameters, i.e. the density of the IGM and the relative velocity. On the other hand, a density higher than $n_{\rm IGM}=5\times{}10^{-5}$ cm$^{-3}$ seems unrealistic for isolated galaxies (but this might be due to the lack of dedicated observations, as the paper by McConnachie et al. 2007 suggests). As a lower density of the IGM would produce smaller perturbations, we infer that ram pressure cannot induce strong lopsidedness.

Another point to stress is the role of the orientation of the relative velocity. We explored different inclinations of the relative velocity vector with respect to the plane of the galaxy, from 0 to 45$^\circ{}$. Large inclinations ($\gtrsim{}20^\circ{}$) produce tidal tails which are non-coplanar with the galaxy, and this does not look like lopsidedness. Only for smaller inclinations the tidal tail resembles a feature of lopsidedness.

Finally, we did not find any way to reproduce the filament observed in NGC~891 with ram pressure. Ram pressure with non-zero inclination with respect to the plane of the galaxy (e.g. $10^\circ{}$, as shown in Fig.~\ref{fig:fig8}) produces some vertical perturbations in the gas density, but these are transient and much smaller than the observed filament in NGC~891.

\subsection{$m=1$ asymmetry}\label{sec:m1}
In order to quantify our results, we derive the value of the Fourier $m=1$ component for the surface density of our gaseous and stellar discs. $m=1$ asymmetries are the best indicator of lopsidedness in stellar discs (Rix \& Zaritsky 1995; Zaritsky \& Rix 1997; B05).
%%%%%%%%%%%%%%%%%%%%%%%%%%%%%%%%%%% FIGURE 9 %%%%%%%%%%%%%%%%%%%%%%%%%%%%%%%%%%
\begin{figure}
\center{{
\epsfig{figure=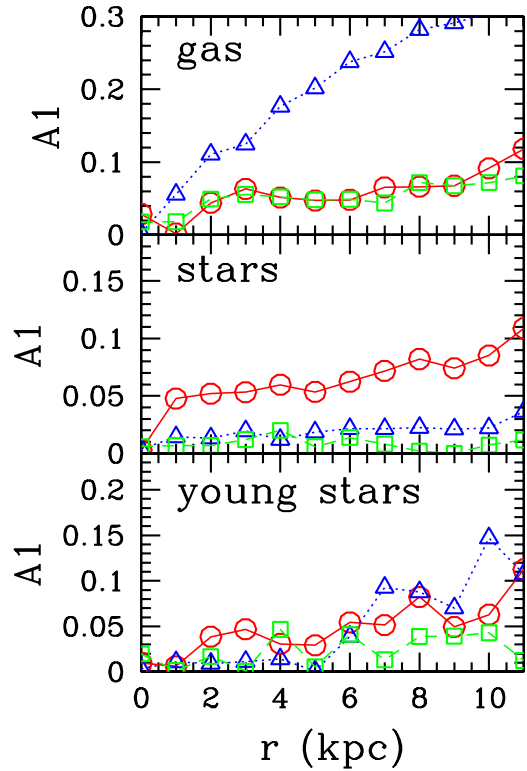,height=11cm}
}}
\caption{\label{fig:fig9} 
Normalized strength of the Fourier component $m=1$ as a function of radius, for  gas (top panel), global stellar content (central panel) and young stellar population (bottom panel). Open circles connected by solid line (red on the web): flyby scenario (at $t=300$ Myr after the interaction); open triangles connected by dotted line (blue on the web): gas accretion (ACC1, at  $t=720$ Myr); open squares connected by dashed line (green on the web): ram pressure (at $t=1$ Gyr).}
\end{figure}
%%%%%%%%%%%%%%%%%%%%%%%%%%%%%%%%%%%%%%%%%%%%%%%%%%%%%%%%%%%%%%%%%%%%%%%%%%%%%%%

Fig.~\ref{fig:fig9} shows the normalized strength of the Fourier component $m=1$ (A1) as a function of radius, for all the considered models. A1 has been defined as in equation~(1) of B05. We calculated A1  for both the stellar and the gaseous component. In the case of the stellar component we distinguish between the global stellar population (central panel of Fig.~\ref{fig:fig9}) and the young stars (bottom panel of Fig.~\ref{fig:fig9}).
%, although only the former is comparable with observations. 
The behaviour of A1 for the gaseous component is shown in  the top panel of Fig.~\ref{fig:fig9}. In the case of gas accretion (ACC1, triangles in Fig.~\ref{fig:fig9}) A1 rises very steeply, and reaches the value of $\sim{}0.3$ at $2-2.5\,{}R_g$. A1 rises also in the flyby (circles in Fig.~\ref{fig:fig9}) and in the ram pressure (squares in Fig.~\ref{fig:fig9}) scenarios, indicating lopsidedness, but it reaches smaller values ($\sim{}0.1$). 

We also calculated the value of $\langle{}{\rm A1}\rangle{}$, i.e. the average value of A1 between 1.5 and 2.5 disc scalelengths (see B05). According to B05, a galaxy is 'lopsided' when $\langle{}{\rm A1}\rangle{}\ge{}0.05$.
For the gaseous component, we found  $\langle{}{\rm A1}\rangle{}\sim{}$0.08, 0.28 and 0.06 for the flyby, gas accretion and ram pressure scenario, respectively.
Thus, all these three scenarios induce lopsidedness in the gaseous component, but the asymmetry produced by gas accretion is much stronger than the one due to flybys and ram pressure. This confirms what we found from a qualitative analysis in the previous sections.

The central panel of Fig.~\ref{fig:fig9} shows the behaviour of A1 for the global stellar component. In the flyby scenario the lopsidedness of the stellar population is similar to the one of the gaseous component: A1 rises almost monotonously from $\sim{}0$ at the centre to $\sim{}0.1$ at $2.5\,{}R_d$. This is due to the fact that the flyby affects both the gaseous and the stellar component. Instead, in the gas accretion and in the ram pressure scenarios A1 remains almost flat through the entire disc. Similarly, for the stellar component, $\langle{}{\rm A1}\rangle{}\sim{}$0.08, 0.02 and 0.01 for the flyby, gas accretion and ram pressure scenario, respectively. This suggests that the stellar population is not lopsided in the case of gas accretion and ram pressure.

This finding might seem inconsistent with the results by B05, where gas accretion is shown to induce lopsidedness also in the stellar component. However, fig.~17 of B05 shows that $\langle{}{\rm A1}\rangle{}\lesssim{}0.05$ for $t<0.7-0.8$ Gyr, in agreement with our results. A value of $\langle{}{\rm A1}\rangle{}\sim{}0.2$ is reached only at $t\sim{}2$ Gyr in the simulations presented by B05. This means that, in the gas accretion scenario, there is a time delay between the epoch in which lopsidedness appears in the gaseous distribution and the epoch in which even the stellar component starts to be lopsided. The extent of this delay depends mainly on the efficiency of star formation.

 This interpretation is confirmed by the bottom panel of Fig.~\ref{fig:fig9}, which shows the behaviour of A1 for the young stellar population ($\le 100$ Myr). In the case of gas accretion the lopsidedness of the young stellar population  is much higher than that of the total stellar population: $\langle{}A1\rangle{}$ is $\sim{}0.09$ and $\sim{}0.02$ for the young stars and for the total stellar mass, respectively. The difference between the distribution of young stars and that of the total stellar component in the gas accretion scenario is also evident, qualitatively, in Fig.~\ref{fig:fig5} (see Section~\ref{sec:accr}). In the case of flybys, the young stars are lopsided ($\langle{}A1\rangle{}\sim{}0.07$) approximately in the same amount as the global stellar content ($\langle{}A1\rangle{}\sim{}0.08$). In the case of ram pressure, the young stars are not lopsided ($\langle{}A1\rangle{}\sim{}0.03$), as already noted for the entire stellar population.

%\subsection{Stellar lopsidedness}
%The analysis of Fig.~\ref{fig:fig9} in Section~\ref{sec:m1} confirms the idea that there are two possible forms
%manifestations 
This supports the idea that there are two possible forms
of stellar lopsidedness: in the first case the lopsidedness of the stars originates at the same epoch and for the same mechanism as the lopsidedness of gas (e.g. for a flyby), in the second case the lopsidedness of the stellar component is a consequence of the (pre-existing) lopsidedness of the gas (e.g. for gas accretion). In the latter case, lopsidedness is present only (or mostly) in the young stellar population, unless the system had enough time to evolve after the process which induced lopsidedness in the gas. For example, after $\sim{}1$ Gyr from the beginning of gas accretion, the difference between spatial distribution of old and young stellar population is large, whereas
no significant difference in the lopsidedness of young and old stars is expected, if the gas accretion from cold filaments started much more than $\sim{}2$ Gyr ago (and cosmological simulations suggest that cold gas accretion was more important at high redshift than today, Keres et al. 2005).

Most of observational work on stellar lopsidedness (Rix \& Zaritsky 1995; Zaritsky \& Rix 1997; Rudnick \& Rix 1998) focuses on quite 'red' bands (especially $I$ and $K$), in order to track the mass distribution. As young stars account only for a small fraction of the light in these bands (especially in early-type disc galaxies), we expect that the observed stellar lopsidedness is different from what we find in the gas accretion scenario, unless gas accretion started sufficiently far ago.
Furthermore, observational studies indicate that lopsidedness in the stellar light distribution is primarily tracing lopsidedness in the total stellar mass distribution (Rudnick \& Rix 1998; Reichard et al. 2008). In particular, the incidence of stellar lopsidedness is similar in late-type ($\sim{}30$ per cent, Rix \& Zaritsky 1995) and in early-type disc galaxies ($\sim{}20$ per cent, Rudnick \& Rix 1998), indicating that the asymmetric light distribution is not due to recent asymmetric star formation, but reflects global asymmetric mass distribution. This implies that the observed stellar lopsidedness may be connected with flyby events or with long-lived gas accretion, but not with recent ($\lesssim{}1$ Gyr) gas accretion. 

On the other hand, Zaritsky \& Rix (1997) and Rudnick et al. (2000) show that there is a correlation between lopsidedness and star formation rate. This suggests that, even if lopsidedness is not due to recent star formation, the event which produced lopsidedness has also triggered star formation. This matches quite well both the flyby and the gas accretion scenario, where a star formation enhancement by a factor of $\gtrsim{}2$ is induced by the galaxy interaction and by the feeding of fresh cold gas, respectively.

%No significant star formation enhancement is associated with the ram pressure case.}

% Thus, observed lopsidedness is generally not due to a small number of 
% As young massive stars account only for a small fraction of the total mass, 
%Furthermore, most of observations focus on relatively 'red' bands ($V-$, $R-$ and $I-$) consider 

%In particular, Rudnick \& Rix (1998) find that the incidence of lopsidedness is similar in early-tipe disc galaxies ($\sim{}20$ per cent) and in late-type ones ($\sim{}30$ per cent, Rix \& Zaritsky 1995). Furthermore, the behaviour of A1 is similar when $V-$, $R-$ and $I-$ bands are considered}
%We stress that most of observational works have focused on I-band or redder imaging (), i.e. they have considered intermediate- and old-stellar populations.

\subsection{Spectrum and kinematics of gas}\label{sec:kine}

%%%%%%%%%%%%%%%%%%%%%%%%%%%%%%%%%%% FIGURE 10 %%%%%%%%%%%%%%%%%%%%%%%%%%%%%%%%%%
\begin{figure}
\center{{
\epsfig{figure=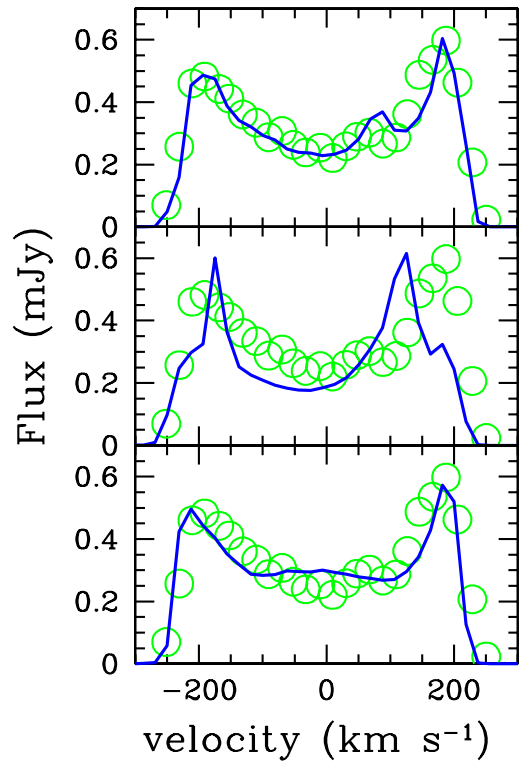,height=11cm}
}}
\caption{\label{fig:fig10} 
HI spectra of the simulated galaxies compared with the observations of NGC~891. Open circles (green on the web): data from Richter \& Sancisi (1994). Solid line (blue on the web): simulations. Top panel: flyby interaction; central panel: gas accretion (ACC1); bottom panel: ram pressure (case reported in Fig.~\ref{fig:fig8}).
}
\end{figure}
%%%%%%%%%%%%%%%%%%%%%%%%%%%%%%%%%%%%%%%%%%%%%%%%%%%%%%%%%%%%%%%%%%%%%%%%%%%%%%%

%%%%%%%%%%%%%%%%%%%%%%%%%%%%%%%%%%% FIGURE 11 %%%%%%%%%%%%%%%%%%%%%%%%%%%%%%%%%
\begin{figure}
\center{{
\epsfig{figure=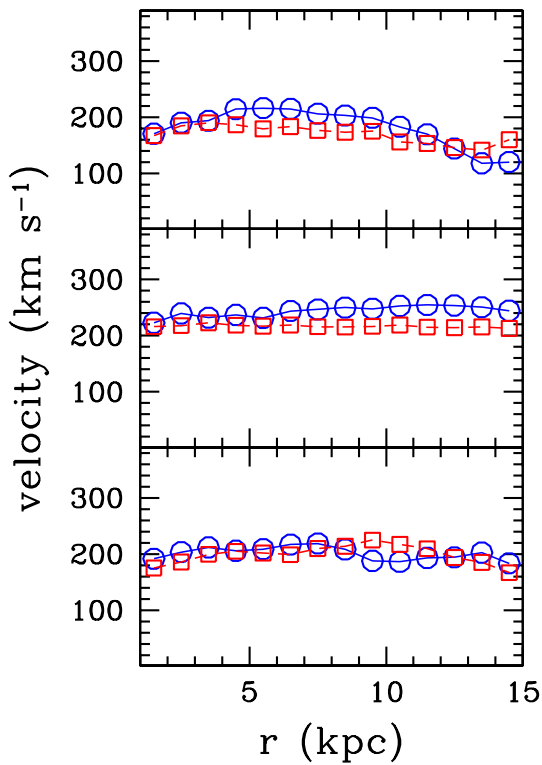,height=11cm}
}}
\caption{\label{fig:fig11} 
Rotation curves of the simulated galaxies for the approaching (open squares connected by dashed line, red on the web) and receding (open circles connected by solid line, blue on the web) sides separately. Top panel: flyby interaction; central panel: gas accretion (ACC1); bottom panel: ram pressure (case reported in Fig.~\ref{fig:fig8}).
}
\end{figure}
%%%%%%%%%%%%%%%%%%%%%%%%%%%%%%%%%%%%%%%%%%%%%%%%%%%%%%%%%%%%%%%%%%%%%%%%%%%%%%%
As we discussed in the previous section, all the considered mechanisms induce a certain degree of lopsidedness in the density distribution of disc galaxies. We now compare the effects that these three mechanisms have on the HI spectrum and on the kinematics, which in the observational data also appear to be affected by lopsidedness. 

In order to derive the HI spectrum and the rotation curve of the galaxy, we adopt a procedure as similar as possible to the one used by observers: we rotate the simulated galaxy by the same inclination angle as the observed NGC~891 (i.e. $\sim{}89^\circ{}$) and we calculate the velocity of gas particles along the line-of-sight (the details of this procedure are given in Mapelli et al. 2008b). 
%By binning the line-of-side-velocity, the number of gas particle in each bin is then approximately proportional to the HI flux. 

The derived HI spectra are shown in Fig.~\ref{fig:fig10} and compared with the observational data of NGC~891 (Richter \& Sancisi 1994). These results are quite interesting, as the gas accretion scenario (central panel) has some problem in reproducing the data, whereas the ram pressure (bottom panel) and especially the flyby scenario (top panel) match quite well the observations.
This might be crucial, if not for all the lopsided galaxies, at least for the case of NGC~891. In fact, other lopsided galaxies have significantly different spectra (see fig.~3 of Richter \& Sancisi 1994), which might be reproduced even by gas accretion (e.g. NGC~5006 or NGC~4303). However, the spectrum of NGC~891 seems to agree only with the scenarios of ram pressure and flyby.

Analogous conclusions can be inferred from the analysis of the rotation curves (Fig.~\ref{fig:fig11}). We calculated two 'partial' rotation curves for each simulated galaxy, one for the receding side of the galaxy (circles in Fig.~\ref{fig:fig11}) and the other for the approaching side (squares in Fig.~\ref{fig:fig11}). This procedure is analogous to the position-velocity diagram and has been adopted to quantify kinematic lopsidedness in Swaters et al.~(1999). 

In the flyby scenario (top panel) the rotation curve declines gently at larger radii than $\sim{}5$ kpc, especially in the receding side. The same trend is apparent from the new data of NGC~891 (see fig.~5 of O07). A similar behaviour can be found in the ram pressure case (bottom panel), although the decrease starts at larger radii and is less pronounced. Instead, in the gas accretion case (central panel) the rotation curve is almost flat everywhere.

Furthermore, in the flyby scenario the rotation curve of the receding side is more steeply rising than the one of the approaching side in the range between $\sim{}5$ and $\sim{}10$~kpc. An analogous trend also exists in the partial rotation curves of NGC~4395 and DD0~9 (Swaters et al. 1999), while the case of NGC~891 is uncertain (O07). Also in the gas accretion scenario the rotation curve of the receding side is steeper, but at larger radii ($\gtrsim{}6-7$ kpc), whereas in the ram pressure case the two sides behave in the same way.
We stress that in some observed (Rubin, Hunter \&{} Ford 1991; Hern\'andez-Toledo et al. 2003) and simulated (Pedrosa et al. 2008) interacting galaxies the receding side of the rotation curve is slightly different from the approaching side. This circumstance may also suggest a connection between lopsidedness and flybys.

\section{Discussion and conclusions}
In this paper we considered three different physical mechanisms acting on disc galaxies: flyby interactions, accretion from cosmological gas filaments and ram pressure from the IGM. We showed that all of these processes might induce lopsidedness  in the gaseous component of discs. Likely, these three scenarios are not exclusive, but might account for lopsidedness in different galaxies. 

Our simulations suggest that the features of lopsidedness induced by these three processes are slightly different from each other. In particular, ram pressure creates only moderate tidal tails in galaxies, whereas flybys account for much stronger asymmetries and gas accretion from filaments might produce even more pronounced lopsidedness. The spectra and the kinematic features connected with these three scenarios are also slightly different. Furthermore, ram pressure is not able to induce lopsidedness in the stellar population, whereas flybys simultaneously produce asymmetries in the gaseous and in the stellar disc. Gas accretion from filaments generates asymmetries only in the gaseous disc, but the star formation in the asymmetric gaseous disc induces lopsidedness in the stellar component at later times. Thus, our simulations suggest that 
%lopsidedness is a variegated process: it may be driven by different mechanisms, and, depending on this, it may involve only the gaseous or also the stellar population, with different amounts and lifetimes.
there is not a unique manifestation of lopsidedness but many different forms: lopsidedness may be driven by various mechanisms, and, depending on this, it may 
%involve only the gaseous or also the stellar population, 
affect only the gas and not (or not yet) the stellar population, it may be more or less pronounced, it may have different timescales. This might also explain why the observed incidence of lopsidedness is different in the stellar and in the gaseous component.
%with different amounts and lifetimes.
Thus, it is interesting to look at the observational properties of lopsided galaxies, and check whether their morphological features allow to infer which process produced their lopsidedness. In particular, the characteristics of  mildly-lopsided  (e.g. NGC~891) and strongly-lopsided (e.g. NGC~1637) galaxies can be studied separately, in order to find any intrinsic difference between them (e.g. differences in the incidence of companion galaxies, in the star formation rate, in the distribution of young stars). Furthermore, lopsided galaxies are also observed in groups and clusters (Haynes, Giovanelli \& Kent 2007, and references therein) where flyby interactions and ram pressure effects can 
be even more important. It would be interesting to compare the fraction of lopsided galaxies as a function of environment.

Another possible feature of lopsidedness which needs to be investigated, both by theoretical models and by observations, are metallicity gradients. 
There are no specific observations of metallicity gradients in lopsided galaxies. 
In the case of both flybys and ram pressure no significant metallicity gradients are expected. Instead, in the case of cold accretion the metallicity of the gaseous filament is likely lower than the one of the target galaxy. 
%For example, Ocvirk et al. (2008) derive a metallicity of $\sim{}0.01-0.03\,{}Z_\odot{}$ for the cold gas accreted at redshift $2.5$, whereas the median metallicity for a Milky Way-like galaxy is $\sim{}0.5\,{}Z_\odot{}$. 
For example, Ocvirk et al. (2008) derive a metallicity  $Z\sim{}10^{-6}-10^{-2}\,{}Z_\odot{}$  (with a very large spread) for the cold gas accreted from low density filaments at redshift $z=4$, whereas the median metallicity for a Milky Way-like galaxy is $\sim{}0.5\,{}Z_\odot{}$. However, it is unclear whether this feature is preserved (and observable) in the nearby Universe. In fact, it is likely that stars form from a mixture between the gas in the cold filament and the gas which was already in the disc. Furthermore, if the gas accretion from the filament is not recent, but occurred in the past, pollution from supernovae, stellar winds and other feedback mechanisms have already erased any metallicity gradient both in the gaseous and in the stellar component.

Here, we focused our analysis on the case of NGC~891. Its morphology, the existence of a companion and of a gaseous filament %between the two galaxies
pointing towards it, the observed HI spectrum and also the shape of the rotation curve  favour the hypothesis of a flyby as source of lopsidedness. 
However, we cannot exclude other scenarios. In particular, the origin of the gaseous filament observed in NGC~891 might be completely different from that of lopsidedness (e.g. a galactic fountain; but see Fraternali \& Binney 2006).
Other observed galaxies (e.g. NGC~2985, Sancisi et al. 2008) show features of lopsidedness similar to those of NGC~891 and have nearby companions. For example,  NGC~3027, the companion of  NGC~2985, has an asymmetric HI distribution. 
%These features suggest that even the lopsidedness of NGC~2985 may be  connected with flyby interactions.

%The role of galaxy interactions in producing lopsidedness has long been discussed, and some previous studies (Wilcots \& Prescott 2004; B05) tend to exclude a link between these processes
Previous studies (Wilcots \& Prescott 2004; B05 and references therein) tend to exclude a link between  galaxy interactions and lopsidedness. However, B05 do not consider flybys but only minor mergers. The main difference between a merger and a flyby is that the merger remnant
remains visible inside the target galaxy for a long time after the beginning of
the merger. Thus, if lopsidedness was due to mergers, we should also
detect the remnants, which is not the case\footnote{However, some lopsided galaxies (e.g. M101) show the signatures of mergers occurred in the past (see Sancisi et al. 2008 and references therein).}.
Instead, in a flyby, the companion
goes away after the interaction. As the effects of the flyby are visible
for at least $\sim{}$500 Myr after the interaction, the companion has enough time to travel far from the target galaxy.

Previous observations do not find a significant correlation between lopsidedness and nearby companions. However, observations such as those by Wilcots \& Prescott (2004) consider
only companions which are still interacting with the target galaxies. In such analysis,
% as Wilcots \& Prescott (2004) point out, 
 companions which had a non-recent flyby with the target are ignored. Even NGC~891 would be considered an isolated galaxy according to this method. Similarly, B05 measure the incidence of companions by the tidal parameter (see equation 4 of B05). The tidal parameter of NGC~891 is $\sim{}-5.0$, which classifies this galaxy as rather isolated (see fig. 9 of B05).

Furthermore, most of the previous studies assume that lopsidedness is produced by the
same process in all the observed galaxies. In this paper we showed that at least three different processes may be responsible for lopsidedness. Thus, the problem of lifetime of
lopsidedness induced by flybys disappears: it may be that long-lived
lopsidedness is connected with gas accretion, and in this
case the lopsided galaxy appears completely isolated; whereas short-lived
lopsidedness is connected with flyby, and in this case the lopsided galaxy
has a companion, although non-necessarily an interacting companion.
%We are not saying that flybys are sources of lopsidedness, but 
Thus, we argue that the role of flybys in producing lopsidedness cannot be excluded on the basis of current observations and simulations.

Therefore, an important issue to address is the rate of flybys which may induce lopsidedness. Here we try to do an approximate calculation. Another class of galaxies is known to be produced by flybys: the collisional ring galaxies (CRGs, see Mapelli et al. 2008a, 2008b and references therein). The density of CRGs in the local Universe is $n_{\rm CRG}\sim{}5.4\times{}10^{-6}\,{}h^3\,{}$Mpc$^{-3}$ (where $h$ is the Hubble parameter, Few \& Madore 1986). Ring galaxies are produced during interactions where the mass of the intruder  is $\ge{}1/20$ of the mass of the target (Appleton \& Struck-Marcell 1996, and references therein). From check simulations we found that lopsidedness is induced during galaxy interactions, if the intruder mass is at least $1/20$ of the target mass. Thus, the mass-ratio between intruder and target is similar (within a factor of 2) for interactions producing ring galaxies and for encounters inducing lopsidedness.
However, other characteristics of flybys which induce lopsidedness are  different from those of interactions which produce CRGs. In particular, 
%the orbit of the intruder is  determined by impact parameter and inclination angle. To 
to produce a ring, the flyby must have a very small impact parameter ($b_{\rm CRG}\lesssim{}0.36\,{}R_d$, corresponding to $\sim{}8$ per cent of the galaxy radius, Few \& Madore 1986). Instead, lopsidedness is produced by off-centre interactions, with impact parameter as large as $b_{\rm max}\sim{}4\,{}R_d$. Furthermore, regular CRGs cannot form when the inclination angle between the disc axis of the target and the intruder is too large ($\sim{}45^\circ{}$, Few \& Madore 1986), whereas flybys producing lopsidedness do not have this limitation. Finally, ring galaxies are short-lived ($t_{\rm CRG}\lesssim{}500$ Myr, Mapelli et al. 2008b), whereas in our simulations the lopsidedness is still present at $t_{\rm lop}\gtrsim{}1$ Gyr after the flyby. 
%Finally, regular CRGs cannot form when the inclination angle between the target and the intruder is too large ($\sim{}45^\circ{}$, Few \& Madore 1986), whereas flybys producing lopsidedness do not have this limitation. 
 Thus, we can derive the density of galaxies which are lopsided as a consequence of flybys ($n_{\rm fly}$) by considering the density of CRGs and by correcting it for these differences\footnote{We remind that the geometric cross-section of interactions with impact parameter smaller than $b$ is proportional to $b^2$.}:
\begin{equation}\label{eq:eqring}
n_{\rm fly}\sim{}n_{\rm CRG}\,{}\frac{b_{\rm max}^2-b_{\rm CRG}^2}{b_{\rm CRG}^2}\,{}\frac{1}{1-\cos{45^\circ{}}}\,{}\frac{t_{\rm lop}}{t_{\rm CRG}}.
\end{equation}
%where $b_1\approx{}R_d$ and $b_2\approx{}3\,{}R_d$ are the minimum and maximum impact parameter to produce lopsidedness (we remind that the cross section of interactions with impact parameter smaller than $b$ is proportional to $b^2$). 
Substituting in equation~(\ref{eq:eqring}) the values reported above, we derive $n_{\rm fly}\sim{}4.5\times{}10^{-3}\,{}h^3$ Mpc$^{-3}$. Since the density of relatively bright disc galaxies (i.e. with absolute magnitude $\lesssim{}-19$ mag) is $n_{\rm disc}\sim{}4.9\times{}10^{-2}\,{}h^3$ Mpc$^{-3}$ (Few \& Madore 1986 and references therein), this suggests that only one galaxy every $\sim{}11$ spiral galaxies can be lopsided as a consequence of flybys. As lopsidedness of the gas component is observed in $\sim{}50$ per cent of disc galaxies (Richter \& Sancisi 1994), flybys can reasonably contribute to $\sim{}20$ per cent of lopsided galaxies. 
%Note that, because $\sim{}30$ per cent of disc galaxies are lopsided in the stellar component (Rix \& Zaritsky 1995; Zaritsky \& Rix 1997), flybys can account to $\sim{}30$ per cent of galaxies whose stellar distribution is lopsided.
%$\sim{}18$ per cent of lopsided galaxies. 

Unfortunately, the density of lopsided galaxies which have undergone gas accretion or ram pressure is much more difficult to estimate. For ram pressure, the peculiar velocities of $\sim{}149$ galaxies in the Local Volume (i.e. $\lesssim{}10$ Mpc from the Milky Way) are known (Whiting 2005) and $\sim{}21$ per cent of them have velocity higher than 100 km s$^{-1}$. However, no data are available for the local density of the IGM, and we cannot even derive an upper limit for the density of ram-pressure perturbed galaxies. For gas accretion, cosmological simulations show that most of galaxies accrete from cosmological filaments. Thus, all galaxies may have undergone cold accretion. Those galaxies in which the accretion was predominantly coplanar may have become lopsided.
%To constrain the fraction of galaxies which accreted cold gas and which became lopsided, new, challenging cosmological simulations are needed.

We stress that accretion from non-coplanar filaments might produce distortions in the disc, different from lopsidedness. This fact deserves further investigations, as it might put constraints on the lifetime, the accretion rate and the existence itself of non-coplanar filaments. Furthermore, the existence of non-coplanar filaments might be associated with the formation of warps (Binney 1992).  

Finally, in this paper, we did not consider two further possible mechanisms for the origin of lopsidedness: the development of asymmetries due to lopsided dark matter halos (Jog 1997, 2002) or to displacements between disc and halo (Levine \& Sparke 1998). These mechanisms might coexist with the three scenarios analyzed in this paper. However,  these mechanisms cannot   produce  the filament observed in NGC~891.

%If lopsidedness in observed galaxies is due to different mechanisms, it would be interesting to compare the morphologies of observed lopsided galaxies, with or without companions, in order to check whether this difference can be found (we suggested the cases of NGC~1637 and NGC~891 as an example).

%-tutti e tre i metodi danno lopsidedness. possono coesistere?

%-per ngc891 flyby

%-halo distorsions non c'entrano con filamento???

%-il filamento puo' essere disconnesso da lopsidedness (galactic fountain). Ma in questo caso la cinematica ancora suggerisce i flybys.

\section*{Acknowledgments}
We thank Renzo Sancisi for his original ideas, vivacious curiosity, helpful discussions and crucial contribution to the paper.
We also thank the anonymous referee for his helpful comments, Simone Callegari for providing the tool to make Fourier analysis, Tobias Kaufmann, Oscar Agertz, Filippo Fraternali, Emanuele Ripamonti and George Lake for useful discussions, and we acknowledge P.~Englmaier for technical support. 
%We also thank Lea Giordano for providing the tools for the Fourier analysis.
MM acknowledges support from the Swiss
National Science Foundation, project number  200020-117969/1
%200020-109581/1
(Computational Cosmology \&{} Astrophysics). J. Bland-Hawthorn is supported  by a
Federation Fellowship from the Australian Research Council.

{}

%\begin{appendix}
%\section{The formation of spokes}
\end{document}